\font\twelverm = cmr10 scaled\magstep1 \font\tenrm = cmr10
       \font\sevenrm = cmr7
\font\twelvei = cmmi10 scaled\magstep1
       \font\teni = cmmi10 \font\seveni = cmmi7
\font\twelveit = cmti10 scaled\magstep1 
       
\font\twelvesy = cmsy10 scaled\magstep1
       \font\tensy = cmsy10 \font\sevensy = cmsy7
\font\twelvebf = cmbx10 scaled\magstep1 \font\tenbf = cmbx10
       \font\sevenbf = cmbx7
\font\twelvesl = cmsl10 scaled\magstep1
\font\twelvett = cmtt10 scaled\magstep1
%
\textfont0 = \twelverm
       \scriptfont0 = \tenrm    \scriptscriptfont0 = \sevenrm
       \def\rm{\fam0 \twelverm}
\textfont1 = \twelvei
       \scriptfont1 = \teni    \scriptscriptfont1 = \seveni
       
\textfont2 = \twelvesy
       \scriptfont2 = \tensy    \scriptscriptfont2 = \sevensy
       
\newfam\itfam \def\it{\fam\itfam \twelveit} \textfont\itfam=\twelveit
\newfam\slfam  \textfont\slfam=\twelvesl
\newfam\bffam \def\bf{\fam\bffam \twelvebf} \textfont\bffam=\twelvebf
       \scriptfont\bffam = \tenbf    \scriptscriptfont\bffam = \sevenbf
\newfam\ttfam  \textfont\ttfam=\twelvett
\rm
\hsize=6.5in
\hoffset=.1in
\vsize=9in
\baselineskip=24pt
%
\raggedright  \pretolerance = 800  \tolerance = 1100
\raggedbottom
%
\dimen1=\baselineskip \multiply\dimen1 by 3 \divide\dimen1 by 4
\dimen2=\dimen1 \divide\dimen2 by 2
%

%
\nopagenumbers
\headline={\ifnum\pageno=1 \hss\thinspace\hss
     \else\hss\folio\hss \fi}
%
\def\heading#1{\vskip \parskip  \vskip \dimen1
     \centerline{#1}
     \vskip \dimen2}
%
\count10 = 0
\def\section#1{\global\advance\count10 by 1
    \vskip \parskip  \vskip \dimen1
    \centerline{\number\count10.\ {#1}}
    \global\count11=0
    \vskip \dimen2}
%
\def\subsection#1{\global\advance\count11 by 1
    \vskip \parskip  \vskip \dimen1
    \centerline{\number\count10.\number\count11.\ {\it #1}}
    \global\count12=0
    \vskip \dimen2}
%
\def\subsubsection#1{\global\advance\count12 by 1
    \vskip \parskip  \vskip \dimen1
    \centerline{\number\count10.\number\count11.\number\count12.\ {\it #1}}
    \vskip \dimen2}
%
%
\def\refindent{\advance\leftskip by 24pt \parindent=-24pt \parskip=2pt}
%
\def\journal#1#2#3#4#5{{\refindent
                      {#1}        
                      {#2},       
                      {#3},       
                      {#4},       
                      {#5}.       
                      \par}}
%

%

%

%

%

%

%

%
\def\figcap#1#2{{\refindent
                   Fig. {#1}.---   
                        {#2}       
                        \par}}
%
\def\etal{{\it et al.\/\ }}
\def\eg{{\it e.g.\/\ }}

\def\lsim{\raise0.3ex\hbox{$<$}\kern-0.75em{\lower0.65ex\hbox{$\sim$}}}
\def\gsim{\raise0.3ex\hbox{$>$}\kern-0.75em{\lower0.65ex\hbox{$\sim$}}}
%
\line{\hfill}
\centerline{\bf A COMPARISON OF COSMOLOGICAL HYDRODYNAMIC CODES}
\vskip 2cm
\centerline{Hyesung Kang$^{1,2}$, Jeremiah P. Ostriker$^1$,  Renyue Cen$^1$,
Dongsu Ryu$^{1,3}$}
\vskip 1cm
\centerline{Lars Hernquist$^4$, August E. Evrard$^5$, Greg L. Bryan$^6$
and Michael L. Norman$^6$}
\vskip 1cm
\centerline{Nov 5, 1993}
\vskip 2cm
\centerline{Submitted to the Astrophysical Journal}
\vskip 1cm
\line{$^1$ Princeton University Observatory, Princeton, New Jersey 08544
\hfill}
\line{$^2$ mailing address: Department of Earth Sciences, \hfill}
\line{\qquad         Pusan National University, Pusan, Korea \hfill}
\line{$^3$ mailing address: Department of Astronomy and Space Science,
\hfill}
\line{\qquad         Chungnam National University, Daejun, Korea\hfill}
\line{$^4$ Board of Studies in Astronomy and Astrophys,
\hfill}
\line{\qquad University of California,Santa Cruz, California.\hfill}
\line{$^5$ Department of Physics, University of Michigan, Ann Arbor,
MI 18109\hfill}
\line{$^6$ National Center for Supercomputing Applications, Urbana, Illinois
61801
\hfill}
\vfill\eject
\heading{ABSTRACT}
We present a detailed comparison of the simulation results of
various cosmological hydrodynamic codes.
Starting with identical initial conditions based on the Cold Dark
Matter scenario for the growth of structure,
with parameters
$h=0.5$,
$\Omega=\Omega_b=1$,
and $\sigma_8=1$,
we integrate from redshift $z=20$ to $z=0$
to determine the physical state within a representative volume of size
$L^3$ where $L=64 h^{-1} {\rm Mpc}$.
Five independent codes are compared: three of them Eulerian
mesh based and two variants of the Smooth Particle Hydrodynamics
"SPH" Lagrangian approach.
The Eulerian codes were run at
$N^3=(32^3,~64^3,~128^3,~{\rm and},~256^3)$ cells,
the SPH codes at $N^3= 32^3$ and $64^3$ particles.
Results were then rebinned to a $16^3$ grid with the expectation
that the rebinned data should converge, by all techniques,
to a common and correct result as $N \rightarrow \infty$.

We find that global averages of various physical quantities do,
as expected, tend to converge in the rebinned model, but that uncertainties
in even primitive quantities such as $\langle T \rangle$,
$\langle \rho^2\rangle^{1/2}$ persists at the 3\%-17\%
level after completion of very large simulations.
The two SPH codes and the two shock capturing Eulerian codes
achieve comparable and satisfactory accuracy
for comparable computer time in their
treatment of the high density, high temperature regions as
measured in the rebinned data;
the variance among the five codes (at highest resolution)
for the mean temperature (as weighted by $\rho^2$)
is only 4.5\%.
Examined at high resolution we suspect
that the density resolution is better in the SPH codes and the
thermal accuracy in low density regions
better in the Eulerian codes.
In the low density, low temperature regions the SPH codes have
poor accuracy due to statistical effects,
and the Jameson code gives temperatures which are too high,
due to overuse of artificial viscosity in these high Mach
number regions.

Overall the comparison allows us to better estimate errors,
it points to ways of improving this current generation
of hydrodynamic codes and of suiting their use to problems
which exploit their individually best features.
\vskip -0.1cm
\noindent{\it subject headings:} cosmology: theory - hydrodynamics - methods:
numerical

\vfill\eject
\section{INTRODUCTION}
Numerical simulations in cosmology have become powerful tools for the
investigation of the quantitative consequences of the various
(and proliferating) scenarios for the growth of structure and also
for relating observational results concerning galaxy distributions
back to more fundamental characterizations of the universe.
Until fairly recently the method of choice was (after adopting
some set of initial conditions) to integrate forwards in time
the equations of motion for a set of collisionless particles in a
fluctuating gravitational field, which itself is
determined self consistently by
the solution of Poisson's equation, given the contemporaneous
density distribution determined by the particle positions.
Examples of this method are the papers of Davis, Efstathiou,
Frenk, and White (in various combinations, \eg 1985a,b),
Melott and colleagues (\eg Centrella \etal 1980),
Bertschinger and colleagues (\eg Bertschinger \& Gelb 1991),
Villumsen and colleagues (\eg Brainerd \& Villumsen 1992)
and others.
The methods are highly developed and many important results
have been obtained.

But, ultimately, when we observe the universe,  we are typically
observing the baryonic component, even if most of the matter
resides in a collisionless "dark" component, which
is treated appropriately by the techniques noted above.
The X-ray emitting gas in clusters of galaxies, the Lyman alpha
clouds, the hot or moving gas responsible for secondary
fluctuations of the microwave background, and the gas from
which galaxies were formed - this material - satisfies the
classic Euler equations of hydrodynamics.
If we wish to model these phenomena, we must solve those
equations as well as the collisionless equation for the
dark matter component (should such exist).

Several techniques have evolved for treating the gaseous
component.  One family, which traces itself back to the
N-body methods familiar to astronomers, is called
"Smooth Particle Hydrodynamics".  Here one follows
representative mass elements by methods analogous to classical
dynamics but attaches thermodynamic variables
(such as pressure) to the particles,
so their motion can be affected by positions
of other nearby particles (pressure gradients), and one has the
possibility of treating, in some approximation, shocks
where fluid elements moving in a given direction overtake
and collide with other fluid elements,
the collision transforming relative fluid velocity into thermal
energy.

This is not the place for a technical exposition concerning methods.
For a detailed presentation one should refer to Hernquist \& Katz (1989)
for the SPH method utilized in this paper and designated TSPH
and refer to Evrard (1988) for the Evrard version which is
designated PSPH hereafter.
The TSPH code uses the Tree method (Barnes \& Hut 1986)
and PSPH uses the P$^3$M (Efstathiou \& Eastwood 1981)
method to calculate the  gravitational field.
The SPH methods are intrinsically Lagrangian in nature in that
they follow individual fluid elements.
This has the advantage that it puts the computational
resources where they are needed most.
Needless to say there are also attendant disadvantages to this
method,
one of which is that the thickness of a shock
discontinuity is typically several times the interparticle
separation.

Classical fluid mechanics, as exemplified by modern aerospace codes,
has followed more typically the Eulerian approach,
where one
uses a fixed (or adaptive) mesh,
which is often, but not necessarily,
Cartesian.  Flux conserving codes have been developed
which integrate the equations efficiently and accurately.
In the recent past shock capturing methods have been
combined with these codes which can stably represent the
density and entropy jumps at shocks within two or three cells.
Two examples of this are
the code written by Ryu \etal (1993) based on
the total variation diminishing scheme (TVD, hereafter)
and the code written by Bryan \etal (1993) based on
Piecewise Parabolic Method (PPM, hereafter).
They have a second-order accuracy in time
in contrast to "COJ" (Cen 1992), which is first order.
Spatially, PPM aims at third order accuracy, TVD second order
and COJ first order.
For given computational resources the Eulerian codes can treat far
more cells than the SPH codes can treat particles, but the
cells are, in the simplest applications, uniformly spaced rather
than concentrated into the more interesting high density regions
as occurs naturally in the SPH codes.

Altogether dozens of papers have been written using these various
methods with various claims made for the accuracy of the methods.
The sceptical reader may well ask what credence should be given
to such claims?
What is the accuracy of these codes?
If given the same initial conditions, do the different methods
arrive at the same results?
If not, which are to be trusted more?
Or, more appropriately, which are to be trusted more in which type
of region?
Finally do they all converge ultimately (at fixed resolution),
with application of sufficient resources, to the same final state,
and if so at what rate?
It is to answer these questions, as objectively as possible,
that this paper is being written.

\section{COMPARISON}
\subsection{Test Calculations}
We have calculated the adiabatic evolution of a purely baryonic
universe but with an initial CDM power spectrum, using the
five cosmological hydro codes described in \S 1.
The purpose of such a test calculation is to focus primarily on the
hydrodynamic properties of these codes (and not, for example,
on their gravity solvers).
The random Gaussian initial conditions have been generated
from the initial power spectrum using the process described
in Weinberg \& Gunn (1990) with the normalization
that the mass fluctuation in a sphere of radius $8h^{-1}{\rm Mpc}$
is $\sigma_8$.
The values of the usual cosmological parameters are
$\Omega = \Omega_b = 1,~h=0.5,$ and $\sigma_8=1$, and the size of the
computational cube is $L=64 h^{-1}{\rm Mpc}$.
The universe has been evolved from $z=20$ to $z=0$.
For Eulerian codes (COJ, TVD, and PPM) simulations with
$32^3,~64^3,~128^3$ and $256^3$ cells were performed.
For SPH codes $32^3$ and $64^3$ particles are used.
We generated the grid data from the SPH particle data by
first smoothing over the interparticle scale and then
integrating the smoothed data over the Eulerian grid cells.

The comparisons are made at the final
epoch, $z=0$.
The results at all these resolutions are rebinned onto a $16^3$ grid in
order to make comparisons of the averaged quantities at a common coarse
resolution.
We have also made some comparisons using the higher resolution original
data.
Hereafter, we will adopt the notation for the simulations in such a way
that TVD128, for example, means TVD $128^3$ simulation
either at full resolution or as rebinned to a $16^3$ grid.
The simulation with the COJ code was carried out by Cen,
the TVD code by Kang, the PPM code by Bryan, the TSPH code by Hernquist
and the PSPH code by Evrard.
The required computational resources used for each code are given
in Table 1.
The hours of computer time listed in column 6 (in units of hours
on a single processor Cray-YMP) are rough estimates.
The listed error is from the Layzer-Irvine cosmological
energy conservation equation and follows the
prescription for $\epsilon= 100|1-R|$ with the latter as defined
by Eq. (3.10) of Ryu \etal (1993).
Note that the $64^3$ SPH codes require comparable or slightly
more time ($\sim 10^2$ hours) than the $256^3$ Eulerian codes
($\sim 10^{1.5}$ hours), but considerably less memory ($10^{-1}-
10^{-1.5}$ as much) to achieve a comparable
({\it cf.}, \S2.2)
overall accuracy for results rebinned to low resolution.

\subsection{Comparison of Rebinned Data on a $16^3$ Grid}
First let us look at integral properties of the results.
We have calculated the average temperature, $\langle T\rangle$, the average
temperature weighted by the density, $\langle T\rangle_{\rho}\equiv
\langle\rho T\rangle/\langle\rho\rangle$,
the average temperature weighted by the density squared,
$\langle T\rangle_{\rho^2}\equiv \langle \rho^2 T\rangle
/\langle\rho^2\rangle$,
and the rms density fluctuation,
$\sigma^2 \equiv (\langle\rho^2\rangle/\langle\rho\rangle^2 -1)$,
on the data rebinned to a $16^3$ grid.
We also calculated the total X-ray luminosity, $ L_x \equiv \Sigma
\rho^2 T^{1/2}$ (in arbitrary units) over the $16^3$ grid.
Because of the coarse smoothing of the density field,
this measure of $L_x$ underestimates by large
factors the true luminosity.
These quantities are summarized in Table 2.
We plot $\langle T\rangle$, $\langle T\rangle_{\rho}$, $L_x$, and
$\sigma$ versus the cube root of the
number of the cells or particles in Fig.~1.
Table 2 and Fig. 1 provide good measures of the rate of convergence
of the simulations in terms of the shocked high density regions,
but they are insensitive to the distribution of the unshocked
low density IGM.
According to Table 2, the PSPH32 and PSPH64 runs, for example, seem to
be converged well with regard to density.
But this does not necessarily mean, for example,
that PSPH64 is more correct in all respects than the other methods,
because the tabulated parameters are not much affected by
the low density region where there are only small number of particles.

In general both SPH code runs have slightly higher temperatures
than the grid-based Eulerian codes.
At low resolution the difference is significant but we see in
Table 2 that the high resolution codes converge
well in the high temperature high density regions.
Specifically the (density)$^2$ average of the temperature is
$(55.3,51.9,58.0,55.0)$
in the highest resolution runs of the
(TVD,PPM,TSPH,PSPH) codes
respectively,
corresponding to a mean of $55.0$ with variance $2.5$ or
only 4.5\%.
With regard to this variable the Eulerian codes appear to
converge from below (they underestimate $\langle T\rangle$)
as does PSPH while for
and
TSPH the situation is less clear.
The Jameson code (COJ) has poorer resolution than the TVD or PPM codes,
since it is a first-order method, while the other two codes are
based on higher order schemes.
We note from Table 1 that it (COJ)
is correspondingly more economical
with regard to both memory and CPU requirements.
The highest level of density fluctuation as rebinned to $16^3$
seems to be achieved in the TVD256 simulation ($\sigma=2.80$),
even though the PPM runs with $N<256$
have generally higher resolution than the comparable
TVD runs (and we will see in 3b and 6c that
in unbinned high resolution results TVD and PPM are almost identical).
The average temperature seems to show a very good convergence for
TVD256 run, while the density and the density weighted temperature
indicate that somewhat more than $256^3$ cells would be
required for the convergence of the density and pressure structure.

A more detailed and definite set of comparison can be made on a
cell-by-cell basis by comparing the
density, temperature and pressure at each cell of the $16^3$ grids
of two different simulations. The density ratios, for example,
are calculated by the density of a "test" run divided by
that of the "control" run at each cell.
The ratios are grouped into bins according to the density of the
"control" run.
Then we calculated the median and first and third quartiles of the
distribution of the logarithm of the ratios in each density bin of
the "control" run.
Fig.~2a shows the median of the logarithm of the density ratios
for each code with TVD256 being picked arbitrarily
as the "control" run.
The displayed error bars are constructed from the differences between
first or third quartiles and the median.

The density distributions of the three Eulerian codes are in a good
agreement as seen in the left hand panels of (2a).
But the density in the SPH codes
in the low density regions tends to be higher than that of the TVD256
and to have larger scatter.
This is true also for unbinned original data and
is probably due to the small number of SPH particles
in the low density regions.
A sign that this may be attributed to error of the SPH code (in this
coarse comparison) is
that the higher resolution TSPH64 run is closer
to the TVD256 run than is the TSPH32 run,
indicating
convergence towards the Eulerian result (in this comparison).
The trend in the low density regions is approximately
reversed if the "test" and "control" runs switch their places.
The interpretation of the high density regions, however, is somewhat
complicated, because the number density of cells with high
density ($\rho/\bar\rho>1$) decreases with the density and so the probability
of spatial overlapping of high density cells between two codes
becomes smaller.
Suppose we take the ratios of density around two high
density clusters which are similar in shape, but the
locations of the clusters in the
two simulations are slightly different.
The ratio is more likely smaller than one for higher density bin,
while it is more likely greater than one for lower density bin,
because of the selection effect of the binning according to the
density of the control run.
This mismatching effect dominates in most of cases in Fig.~2a
except perhaps $\rho(COJ256)/\rho(TVD256)$ versus $\rho(TVD256)$.

Another, more impartial way to look at the high density regions is to
plot a scatter diagram of ${\rm log}(\rho_1/\rho_2)$ versus
${\rm log}(\rho_1\rho_2)^{1/2}$ as in Fig.~2b where
$\rho_2=\rho(TVD256)$.
Except for $\rho(COJ256)/\rho(TVD256)$ and $\rho(TVD128)/\rho(TVD256)$
(which have lowest resolution),
the points are randomly scattered about
the mean, indicating that there is no significant
difference in the high density regions among runs with different codes
as rebinned to $16^3$ resolution.

Now let us turn to the temperature variable, which is most sensitive
to details of the hydrodynamic treatment.
The ratios of the temperature versus the temperature of TVD256
are plotted in Fig.~2c.
For all cases, the temperature for high temperature regions seems
to agree fairly well among the various simulations with
the agreement beween the two high resolution codes (TVD, PPM:
bottom left panel) quite good.
As noted, the SPH values of temperature
are slightly higher than the Eulerian values
in the high density regions.
The temperature of COJ256 for low temperature regions is higher than
that of TVD256,
because the COJ code has more artificial viscosity in low
temperature regions with supersonic bulk motion.
On the other hand, the temperature of the PPM run for low
temperature regions
is somewhat lower than that of TVD256.

The similar plots for the temperature ratios versus
the gas density of TVD256 are shown in Fig.~2d.
This shows that the temperature of high density regions is quite
similar for all codes, but that of low density region displays
significant differences. These results are consistent with Fig.~2c, since
the high density cells also tend to have a high temperature.

Figs.~3a-b show the volume and mass weighted histograms of the
density in $16^3$ cells for TVD256, PPM256, TSPH64, and PSPH64.
The TVD256 and PPM256 runs have somewhat broader distributions than
the SPH runs, because they have
more lower density cells and more higher density cells.
The agreement between the two Eulerian codes
and between the two SPH codes is
excellent.
Fig.~3c-d shows the volume and mass weighted histograms of the
temperature for the same cases as in Fig.~3a,b.
Both SPH codes have somewhat more hot gas than TVD256 or PPM256.
As expected, the volume occupied by the high temperature
regions with $T\gsim10^6 {\rm K}$ are similar in PPM and TVD codes.

Figs.~4a and 4b show the density and temperature contour plots
of a slice with $16 h^{-1}{\rm Mpc}$ thickness on $16^3$ grids in each
the simulation.
The density was normalized by the mean baryon density,
while the temperature
was normalized by $T_b=10^6K$ before contouring.
The density contours show that the density distributions on a $16^3$
grid are, in general, quite similar for all codes.
The temperature contours, on the contrary, show some differences.
The SPH runs show more hot gas than the Eulerian code runs
and it is spread out over larger regions perhaps
due to less sharp shocks in low density regions.
The temperature in COJ256 run is lower than than in TVD256 or PPM256,
which is true even when compared with TVD128 or PPM128.
This illustrates that, while the density structure seems to agree
well, the temperature distributions
shows much bigger differences among
various hydro codes even after averaging the data onto a very coarse
grid.

\subsection{Comparison of the Original Data}
Figs.~5a-c show the density and temperature contour plots
of a slice with $0.25 h^{-1}{\rm Mpc}$ thickness (1 cell) on $256^3$
grids for three Eulerian codes, while Fig.~5d shows the similar contour
plots of a slice with $0.32 h^{-1}{\rm Mpc}$ for PSPH64 data mapped onto
$200^3$ cells;
and Figure (5e) shows
a slice of $0.5h^{-1}$Mpc smoothed onto a $256^3$
grid for the TSPH code result.
The temperature contour plot for the COJ256 run indicates the artificial
heating of the low density IGM.
The agreement between the two high resolution
Eulerian codes [(PPM256, TVD256)=(5b,5c)]
is excellent.
One sees in the temperature plots caustics emanating out
of cluster like regions bounded by sharp shocks.
{}From the thickness of the pancakes and the filaments (compared
to the cell size), it is clear that
the detailed structure has not converged even in $256^3$ run.

In Figs.~6a-e
the contour plots of the cells (for Eulerian codes)
or particles (for SPH codes)
with given temperature
and density are shown for three Eulerian and two
SPH codes.  Cells are weighted by mass in the Eulerian codes.
The regions with the high temperature ($T\gsim10^6{\rm K}$) are
quite similar with highest density and temperature
reached in the SPH codes.
As has been seen in the histograms in Figs.~4, the warm and cold gas
regions do not agree well among the four codes.

We calculated the fraction of the mass with given temperature
and density for the TVD256 and PPM256
original data. The same fraction was
calculated for the PSPH64 and TSPH64,
using the original particle density and
temperature, and counting particle numbers instead of mapping the
particles onto a grid.
They are plotted in Figs.~7b with the analogous Eulerian histograms
shown in 7a.
{}From these presentations, we can see that particle density in the
TSPH32 and PSPH64
can become much larger than the grid density in the TVD256 in the
high density postshock regions.
On the other hand, the particle temperature distribution
in the TSPH64 and PSPH64 is
relatively close to that in the TVD256 and PPM256.

The different trends between the density and temperature can
be understood simply, because the temperature is a conserved
quantity (\eg thermal energy per mass) within a particle,
while the density is not (\eg mass per volume).
One of the obvious results is that the X-ray emissivity of the
TSPH64 and PSPH64 particle data will be
much greater than the TVD256, while the X-ray emissivity on
$16^3$ resolution, $L_x$, in Table 2 of the three codes are quite close.
They also show dramatically that the density is overestimated in
the low density regions
of the SPH calculation ($0.2<\rho/\bar\rho <5$ in Fig. 8).
In the SPH results there are no particles with
$\rho/\bar\rho<0.2$, while TVD256 has cells with density
as low as 0.01.
One should note that, with the same number of particles,
the TSPH64 run seems to achieve a slightly
higher resolution than PSPH64.
It is likely that the difference comes from the way with which
the smoothing length varies in time and space in the two codes.

Finally we plotted the density and temperature distribution along
an arbitrary line of sight for the TVD256 and PSPH64 in Figs. 8.
The particles within a cylinder with a square face of $(1 h^{-1} Mpc)^2$
along the line of sight are chosen for the PSPH64 and plotted as open
circles.
Among 16 rows of cells of TVD256 within the same square-faced cylinder,
two at the corner and two near the center are shown.
We can see the density of clusters in PSPH64 can become much higher
than in TVD256, while the temperatures of clusters in the
two runs are very
similar.
Once again, the X-ray appearance of the clusters in the two codes
will be very different.
This figure also illustrates well that only a small number of
particles are in
the low density region to represent the flow in the SPH code,
so the density is overestimated in that region.
Fig.~8b shows a blow-up of a single density peak around $R=33 h^{-1}
{\rm Mpc}$ to examine in close-up the details of the high density cluster.
There could be several explanations for the stronger clustering in
PSPH64 than in TVD256 two of which are the following:
1) it could come from the difference between P$^3$M code and Eulerian
code
for the gravitational field calculation since the SPH codes
clearly have higher force resolution in dense regions;
2) it is possible that PSPH64 achieves hydrodynamically higher
resolution than TVD256 due to its Lagrangian nature.
Further investigation is necessary to discriminate among
these possibilities.

\section{IMPLICATIONS FOR FUTURE WORKS}
It is reassuring to note in Fig.~1 that the various codes do
in fact converge to the same results as $N \rightarrow \infty$.
It appears that all codes converge from below with regard to
density ($\langle \rho^2\rangle / \langle \rho \rangle$ is always
underestimated),
but that with regard to the mean temperature
the situation is not as clear.
The
Eulerian codes appear to converge from below
as does PSPH
while TSPH code may converge from above
(or may not do so, but seem to in Table 1 due to
our smoothing and rebinning procedures).
With regard to the integral quantities, derived from rebinned data,
it appears that $64^3$ SPH results are of an accuracy intermediate
between $128^3$ and $256^3$ in the shock capturing Eulerian codes.
This indicates quite crudely (after examining Table 1) that
by these measures comparable accuracy is reached with comparable
computer time.
The overall accuracy, as measured by convergence,
is perhaps disappointing.
Fig.~1 indicates roughly that some integrated quantities are still
insecure to 20\% in rebinned data (on a $16^3$ grid) for
$N=64^3$ SPH runs and the comparably costly Eulerian codes
whereas other quantities have converged to 5\%.

Overall, the agreement in density, if one compares on a cell by
cell basis, is far better than the agreement in temperature,
especially in the high density regions.
The SPH codes have poor information in the low density regions
and poorer resolution of shocks but much better resolution in high
density regions for comparable expenditure of computer resources.
Agreement of temperature among the simulations in the high
temperature regions is fairly good (given the caveats mentioned
previously).
In the low density ($\rho/\bar\rho <1$), low temperature regions
it appears that there is a quite significant overestimate of
temperature by the COJ code.

One interesting fact is that in the binned data the distributions
of temperature and density are broader in the Eulerian codes
than in the SPH codes but in the unbinned data the converse is true.
It is likely but not certain that the SPH codes give a better
representation of the highest density regions due to their
intrinsically Lagrangian nature, while the Eulerian codes are
probably better in the regions $\rho/\bar\rho < 1$ due to
better sampling.
The temperatures in regions like the X-ray clusters can be
obtained with moderate accuracy with any of the highly
developed codes,
but there is still great uncertainty
in computing the thermal energy of the lower temperature regions,
This could be a problem in cosmological applications
because, if the box scale is chosen to
correspond to a wavelength bigger than that at the peak
of the power spectrum,
the ``low" temperatures correspond to those in regions where
galaxy formation is thought to occur.

These insights should lead us towards future code improvements,
to better estimate of our accuracy and towards applications
of the individual codes to problems for which they are best suited.

\heading{Acknowledgments}
The work by HK, JPO, RC, and DR was supported in part by
NASA through grant NAGW-2448 and by NSF through grants AST91-08103
and ASC-9318185.
HK was supported in part by the Korea Research Foundation through
the Brain Pool Program.
LH was supported in part by NASA through grant NAGW-2422
and by NSF through Presidential Faculty Fellowship.
AEE was supported in part by NASA through grant NAGW-2367.
GLB and MLN was supported in part by NASA through NAGW-3152.
GLB would like to thank Jim Stone for useful discussions.
In the later stages of this work the Princeton, UCSC and NCSA
work was supported by the NSF, HPCC grant ASC93-18185.
\vfill\eject
\centerline{TABLE 1}
\bigskip
\centerline{Required Computational Resources}
\bigskip
\vbox{\tabskip=0pt\offinterlineskip
\def\tablerule{\noalign{\hrule}}
\halign to500pt{\strut#&\vrule#\tabskip=1em plus2em&
\hfil#&#&\hfil#\hfil&#&
\hfil#&#&
\hfil#\hfil&#&\hfil#\hfil&#&
\hfil#\hfil&#&
\hfil#&\vrule#\tabskip=0pt\cr\tablerule
&&\omit\hidewidth${\rm run}$\hidewidth&&
  \omit\hidewidth${\rm Memory^a}$\hidewidth&&
  \omit\hidewidth${\rm CPU^b (machine)}$\hidewidth&&
  \omit\hidewidth${\rm CPU^c }$\hidewidth&&
  \omit\hidewidth${\rm steps}$\hidewidth&&
  \omit\hidewidth${\rm total CPU^d}$\hidewidth&&
  \omit\hidewidth$\epsilon(\%)$\hidewidth&\cr\tablerule
&&TVD128&&34 &&590 (Convex 220)&&59&&130&&2.1&&4&\cr
&&TVD256&&250 &&4390 (Convex 3440)&&450&&153&&19.1&&5&\cr\tablerule
&&PPM128&&32 &&180 (Convex 3880)&&90&&200&&7.2&&9&\cr
&&PPM256&&190 &&169 (64-node CM5)&&559&&255&&57.6&&4&\cr\tablerule
&&COJ128&&16 &&103 (Convex 220)&&14&&490&&1.9&&2&\cr
&&COJ256&&128 &&1000 (Convex 3440)&&100&&500&&13.9&&1&\cr\tablerule
&&TSPH32&&5 &&45 (Cray YMP)&&45&&500&&6.26&&4&\cr
&&TSPH64&&40 &&360 (Cray YMP)&&360&&500&&50.0&&3&\cr\tablerule
&&PSPH32&&0.94 &&240 (Stardent Titan 3020)&&12&&1000&&3.3&&5&\cr
&&PSPH64&&7.5 &&73.8 (HP 735)&&144&&1000&&40&&3&\cr\tablerule }}
\bigskip
\line{ $^a$ in units of Megawords\hfill}
\line{ $^b$ CPU time in units of seconds per time step\hfill}
\line{ $^c$ equivalent CPU time on a single processor Cray-YMP in units of
seconds \hfill}
\line{\quad per time step\hfill}
\line{ $^d$ equivalent total CPU time on a single processor Cray-YMP in units
of hour\hfill}
\vfill
\eject
\centerline{TABLE 2}
\bigskip
\centerline{Integrated Numbers}
\bigskip
\vbox{\tabskip=0pt\offinterlineskip
\def\tablerule{\noalign{\hrule}}
\halign to500pt{\strut#&\vrule#\tabskip=1em plus2em&
\hfil#&#&\hfil#\hfil&#&
\hfil#&#&
\hfil#\hfil&#&\hfil#\hfil&#&
\hfil#&\vrule#\tabskip=0pt\cr\tablerule
&&\omit\hidewidth${\rm run}$\hidewidth&&
  \omit\hidewidth$<T>^a$\hidewidth&&
  \omit\hidewidth$<T>_{\rho}^a$\hidewidth&&
  \omit\hidewidth$<T>_{\rho^2}^a$\hidewidth&&
  \omit\hidewidth$\sigma$\hidewidth&&
  \omit\hidewidth$L_x$\hidewidth&\cr\tablerule
&&TVD32&&$0.720$&&$3.71$&&$13.1$&&$1.52$&&$0.419$&\cr
&&TVD64&&$1.53$&&$9.79$&&$30.2$&&$2.23$&&$1.20$&\cr
&&TVD128&&$1.83$&&$13.1$&&$42.5$&&$2.58$&&$1.86$&\cr
&&TVD256&&$1.85$&&$16.2$&&$55.3$&&$2.80$&&$2.45$&\cr\tablerule
&&PPM32&&$1.05$&&$7.31$&&$24.1$&&$2.15$&&$0.984$&\cr
&&PPM64&&$1.59$&&$11.3$&&$34.4$&&$2.38$&&$1.44$&\cr
&&PPM128&&$1.84$&&$15.1$&&$44.9$&&$2.58$&&$1.91$&\cr
&&PPM256&&$2.01$&&$17.7$&&$51.9$&&$2.70$&&$2.24$&\cr\tablerule
&&COJ32&&$0.519$&&$2.26$&&$6.56$&&$1.52$&&$0.302$&\cr
&&COJ64&&$1.19$&&$6.21$&&$18.2$&&$1.93$&&$0.733$&\cr
&&COJ128&&$1.44$&&$8.86$&&$26.7$&&$2.25$&&$1.16$&\cr
&&COJ256&&$1.57$&&$11.9$&&$38.2$&&$2.49$&&$1.65$&\cr\tablerule
&&TSPH32&&$2.92$&&$24.4$&&$68.4$&&$2.70$&&$2.61$&\cr
&&TSPH64&&$2.52$&&$19.8$&&$58.0$&&$2.65$&&$2.30$&\cr\tablerule
&&PSPH32&&$2.37$&&$18.0$&&$54.5$&&$2.54$&&$2.08$&\cr
&&PSPH64&&$2.46$&&$19.2$&&$55.0$&&$2.67$&&$2.27$&\cr\tablerule}}

\bigskip
\line{$^a$ in units of $10^6K$.\hfill}
\vfill
\eject
\heading{REFERENCES}
\journal{Barns, J., \& Hut, P.}{1986}{Nature}{324}{446}
\journal{Bertschinger, E., \& Gelb, J.}{1991}{Comput.~in Phys.}{5}{164}
\journal{Brainerd, T., \& Villumsen, J.}{1992}{ApJ}{394}{409}
\journal{Bryan, G.~L., Norman, M.~L., Stone, J.~M., Cen, R.,
\& Ostriker, J.~P.}{1993}{Computer Physics Communications}
{}{in preparation}
\journal{Cen, R.}{1992}{ApJS}{78}{341}
\journal{Centrella, J.~M., Gallagher, J.~S., Melott, A.~L., \&
Bushouse, H.~A.}{1988}{ApJ}{333}{24}
\journal{Davis, M., Efstathiou, G., Frenk, C.~S., \& White, S.~D.~M.}
{1985a}{ApJ}{292}{371}
\journal{Efstathiou, G., Davis, M., Frenk, C.~S., \& White, S.~D.~M.}
{1985b}{ApJS}{57}{241}
\journal{Efstathiou, G., \& Eastwood, J.W.}{1981}{MNRAS}{194}{503}
\journal{Evrard, A.~E.}{1988}{MNRAS}{235}{911}
\journal{Hernquist, L. \& Katz, N.~S.}{1989}{ApJS}{64}{715}
\journal{Ryu, D., Ostriker, J.~P., Kang, H., \& Cen, R.}{1993}{ApJ}
{414}{1}
\journal{Weinberg, D.~H., \& Gunn, J.~E.}{1990}{MNRAS}{247}{260}
\vfill\eject
\heading{FIGURE CAPTIONS}
\figcap{1}
{Convergence of global averages.
Average temperature (top left), average temperature weighted
by the density (top right), the total X-ray luminosity (bottom left),
and the rms density fluctuation
($\sigma^2\equiv \langle\rho^2\rangle/\langle\rho\rangle^2-1$)
of the rebinned data on a $16^3$ grid
for the COJ (filled square),
PPM (filled triangle), TVD (filled circle), PSPH (open circle),
and TSPH (open triangle) codes as a function of the cube root of the
number of cells or particles. }

\figcap{2a}{${\rm log}(\rho_1/\rho_2)$ versus ${\rm log}(\rho_2)$
where $\rho_2=\rho({\rm TVD256})$.
The simulation designations for $\rho_1$ are expressed in the labels
on the abscissa.
The solid line is the median of the logarithm of the density ratio
distribution at each bin, while the error bars are constructed from
the differences between first and third quartiles and the median.
Bottom left panel shows that high resolution Eulerian codes
agree well with one another, and comparison of two lower right hand
panels shows that the two SPH codes agree well with each other.
}

\figcap{2b}{${\rm log}(\rho_1/\rho_2)$ versus ${\rm log}(\rho_1\rho_2)$
for high density cells
on a $16^3$ grid. For all cases $\rho_2=\rho(TVD256)$
as in the Fig.~2a.
The simulation designations for $\rho_1$ are expressed in the labels
on the abscissa.  No trends are seen, indicating that
in the rebinned data all codes have similar resolution except COJ256 and TVD128
which reach systematically lower densities in high density
regions.}

\figcap{2c}{Same as Fig.~2a except the temperature ratios
are plotted for the temperature of the TVD256 run.
Note agreement in the high temperature regions
and variance in the low temperature regions.}

\figcap{2d}{Same as Fig.~2c except the temperature ratios
are plotted for the density of the TVD256 run.
Note agreement in the high density regions and variance in
low density regions.}

\figcap{3a}{Mass weighted density histogram for
the TVD256 (top panel, solid line), PPM256 (top panel, dashed line),
TSPH32 (bottom panel, solid line) and PSPH64 (bottom panel, dashed
line) runs. The simulation results at $z=0$ rebinned onto a $16^3$
grid are shown. Eulerian codes show
a slightly broader distribution.}

\figcap{3b}{Volume weighted density histogram for
the same simulations as in Fig.~3a.}

\figcap{3c}{Mass weighted temperature histogram for
the same simulations as in Fig.~3a. SPH codes reach somewhat higher
temperatures.}

\figcap{3d}{Volume weighted temperature histogram for
the same simulations as in Fig.~3a.
Peaks in very low temperature regions are spurious, numerical artifacts.
}

\figcap{4a}{Density contour plots of a slice with $16 h^{-1}{\rm Mpc}$
thickness on a $16^3$ grid for each simulation. General agreement
is seen.}

\figcap{4b}{Same as Fig.~4a except temperature contour plots are
shown. SPH codes show broader peaks possibly due to poorer resolution
of shocks.}

\figcap{5a}{Gas density (top) and temperature (bottom) contour plots
of a slice with $0.25 h^{-1}{\rm Mpc}$ thickness on a $256^3$ grid for
the COJ256 run. Note in (5a-5d) the familiar pattern of caustics
connecting vertices of high temperature gas.}

\figcap{5b}{Same as Fig.~6a except for the PPM256 run.}

\figcap{5c}{Same as Fig.~6a except for the TVD256 run.}

\figcap{5d}{Gas density (top) and temperature (bottom) contour plots
of a slice with $0.32 h^{-1}{\rm Mpc}$ thickness on a $200^3$ grid for
the PSPH64 run.
The location of the slice is the same as the ones in Fig. 5c.}

\figcap{5e}{Gas density (top) and temperature (bottom) contour plots
of a slice with $0.5 h^{-1}{\rm Mpc}$ thickness on a $256^3$ grid for
the TSPH64 run.
The location of the slice is the same as the ones in Fig. 5c}

\figcap{6}{Contour plot of the volume fraction with given temperature
and density. The simulation results at $z=0$ on a $128^3$ grid are
shown.  The plots are for
(a) COJ256
(b) PPM256, (c) TVD128,
(d) PSHP64, and (e) TSPH64. SPH codes reach to higher
temperatures and densities. See text for binning/smoothing parameters.}

\figcap{7a}{Histograms of the mass fraction with given temperature
and density for TVD256 (solid line) and
PPM256 (dashed line) original data.}

\figcap{7b}{Histograms of the mass fraction with given temperature
and density for TSPH64 (dashed line) and
PSPH64 (solid line) original data.}

\figcap{8a}{Density and temperature distribution along an arbitrary
line of sight for the TVD256 original data (left two panels) and
the PSPH64 original particle data (right two panels).
Four different rows of cells in the TVD256 run are shown within a
cylinder with a square face of $(1 h^{-1} {\rm Mpc})^2$,
while the particles
of the PSPH64 run located in the same cylinder are plotted.
}

\figcap{8b}{Same as  Fig.~9a except that blow up of a single peak is shown.}
\vfill
\eject
\bye
\end